\documentclass[superscriptaddress,twocolumn,showpacs,
amssymb,amsmath,nobibnotes,aps,prd,longbibliography,
nofootinbib]{revtex4-1}
\usepackage{graphicx,subfigure,bm,color,psfrag,hyperref}
\usepackage{float}
\usepackage{epsf}
\usepackage{bm}
\usepackage{booktabs}
\usepackage{amsmath}
\usepackage{amsfonts}
\usepackage{amssymb}
\usepackage{epstopdf}
\usepackage{natbib}
\usepackage[normalem]{ulem}
\usepackage[dvipsnames]{xcolor}
\usepackage{array}
\providecommand{\U}[1]{\protect\rule{.1in}{.1in}}
\newcommand{\be}{\begin{equation}}
\newcommand{\ee}{\end{equation}}

\newcommand{\mincir}{\raise
-3.truept\hbox{\rlap{\hbox{$\sim$}}\raise4.truept\hbox{$<$}\ }}
\newcommand{\magcir}{\raise
-3.truept\hbox{\rlap{\hbox{$\sim$}}\raise4.truept\hbox{$>$}\ }}

\definecolor{blue}{rgb}{0.36, 0.54, 0.66}
\definecolor{amaranth}{rgb}{0.9, 0.17, 0.31}
\definecolor{pink}{rgb}{0.87, 0.56, 0.81}
\definecolor{ao}{rgb}{0.0, 0.5, 0.0}
\definecolor{maroon}{rgb}{0.76, 0.13, 0.28}
\definecolor{cardinal}{rgb}{0.77, 0.12, 0.23}
\definecolor{lightcardinal}{rgb}{0.97, 0.42, 0.53}
\definecolor{frenchlila}{rgb}{0.53, 0.38, 0.56}
\definecolor{yellow}{rgb}{1.0, 1.0, 0.87}
\definecolor{lightseagreen}{rgb}{0.7, 0.92, 0.68}
\definecolor{gray}{rgb}{0.9, 0.9, 0.9}
\definecolor{lightblue}{rgb}{0.66, 0.84, 0.96}

\hypersetup{colorlinks,linkcolor={maroon},citecolor={magenta},urlcolor={ao}}

\begin{document}

\title{A note on the gravitational  dark matter production}

\author{Jaume de Haro}
\email{jaime.haro@upc.edu}
\affiliation{Departament de Matem\`atiques, Universitat Polit\`ecnica de Catalunya, Diagonal 647, 08028 Barcelona, Spain}

\author{Supriya Pan}
\email{supriya.maths@presiuniv.ac.in}
\affiliation{Department of Mathematics, Presidency University, 86/1 College Street,  Kolkata 700073, India}
\affiliation{Institute of Systems Science, Durban University of Technology, PO Box 1334, Durban 4000, Republic of South Africa}

\begin{abstract}
Dark matter, one of the  fundamental components of the universe, has remained mysterious in modern cosmology and particle physics, and hence, this field is of utmost importance at present moment. One of the foundational questions in this direction is the origin of dark matter which directly links with its creation. 
In the present article we study the gravitational production of dark matter in two distinct contexts: firstly, when reheating occurs through the gravitational particle production, and secondly, when it is driven by the inflaton's decay. We establish a connection between the reheating temperature and the mass of dark matter, and from the reheating bounds, we determine the range of viable dark matter mass values. 
\end{abstract}

\vspace{0.5cm}


\maketitle

\thispagestyle{empty}

\section{Introduction}

Dark matter, a fundamental component of the universe, remains one of the most profound mysteries in modern cosmology and particle physics. Despite its gravitational effects being observed across a variety of astrophysical scales, from galaxies to the cosmic microwave background, its origin and nature continue to elude us. Among the proposed mechanisms for dark matter generation,  gravitational production of dark matter stands out as a particularly compelling explanation,  especially within the context of the early universe,  see for instance \cite{Chung:1998bt,Chung:2001cb,Markkanen:2015xuw,Ema:2018ucl,Ema:2019yrd,Haro:2019ndy,Haro:2019umj,Cembranos:2019qlm,Cembranos:2020,Babichev:2020yeo,Karam:2020rpa,Mambrini:2021zpp,Bernal:2020ili,Garcia:2022vwm,Bastero-Gil:2023htv,Zhang:2023hjk,Barman:2023icn,Bastero-Gil:2023mxm,Belfiglio:2024xqt,Bertuzzo:2024fns,
Wang:2024lva}.

Gravitational dark matter production leverages the unique role of gravity, a universal interaction, as the primary mechanism for generating dark matter particles. This process requires no additional couplings or interactions with the Standard Model, relying solely on the dynamics of the expanding universe. Such production is especially relevant during the reheating phase following cosmic inflation, where the universe transits from an early inflationary epoch to a radiation-dominated era. 

Two primary scenarios dominate discussions of gravitational dark matter production: one in which reheating occurs through the copious production of heavy particles \cite{Hashiba:2018iff}
that subsequently decay into Standard Model particles, and another where reheating results from the decay of the inflaton field directly into Standard Model particles
\cite{Garcia:2020wiy}. These pathways connect the physics of inflation, reheating, and dark matter, establishing a relationship between the reheating temperature and the mass of dark matter. This connection allows us, through reheating constraints, to identify the range of viable dark matter masses.

This short note delves into the theoretical foundations of gravitational dark matter production
studied in several works \cite{Chung:1998bt,Chung:2001cb,Markkanen:2015xuw,Ema:2018ucl,Ema:2019yrd,Haro:2019ndy,Haro:2019umj,Cembranos:2019qlm,Cembranos:2020,Babichev:2020yeo,Karam:2020rpa,Bernal:2020ili,Garcia:2022vwm,Bastero-Gil:2023htv,Zhang:2023hjk,Barman:2023icn,Bastero-Gil:2023mxm,Belfiglio:2024xqt,Bertuzzo:2024fns,
Wang:2024lva} 
emphasizing its dependence on inflationary reheating dynamics, its potential observational implications, and the range of viable dark matter masses within this framework. This short note has been organized as follows. In section~\ref{sec-2} we present the gravitational reheating formulas. Section~\ref{sec-3} describes the gravitational production of dark matter in the context of gravitational reheating. In section~\ref{sec-4} we discuss the gravitational production of dark matter and reheating through the decay of the inflaton field. Finally, in section~\ref{sec-5-conclusion} we conclude the present note with a brief summary. 

Throughout the article, we work under the assumption of the spatially flat Friedmann-Lema\^{i}tre-Robertson-Walker (FLRW) geometry with $a(t)$ (hereafter $a$) representing the expansion scale factor of the universe  and  
we have used the following notations:
\begin{enumerate}
\item ``END'': denotes the end of inflation.
\item ``$0$'': denotes the present time.
\item ``reh'': denotes the reheating time. 
\item
$ \rho_{A, \rm END}$: denotes the energy density of the produced $A$-particles,  at the end of inflation. 
\item $\rho_{\rm B,\rm END}=3M_{\rm pl}^2H_{\rm END}^2$
is the energy density of the background at the end of inflation  ($M_{\rm pl}$ is the reduced Planck mass), that is, it corresponds to the energy density of the inflaton field.

\item $\rho_r$ corresponds to the energy density of the radiation.

\item $\Theta_A=\frac{\rho_{A,\rm END}}{\rho_{\rm B, \rm END}}$ is the heating efficiency of the $A$-particles. 

\item $\bar{\Theta}_A=\frac{3\Gamma_A^2M_{\rm pl}^2}{\rho_{\rm B,\rm END}}$ is the decay efficiency of the $A$-particles, where $\Gamma_A$ is the decay rate of the $A$-particles.

\item $\Omega_r h^2$: density parameter for radiation  ($h = H_0/100$ km/s/Mpc in which $H_0$ is the present day value of the Hubble constant).

\item $\Omega_A h^2$: density parameter of the $A$-particles.
\end{enumerate}

Concerning the observational constraints on some of the parameters, we have used the following values~\cite{Lahav:2022poa}:
\begin{enumerate}
\item $\Omega_Y h^2=0.12\pm 0.0012$, where the $Y$-particles are the candidate for dark matter.

\item $\Omega_r h^2\cong 2.47\times 10^{-5}$.

\item $h=0.674\pm 0.005$.

\item $T_0=2.7255\pm 0.006 K\cong 2.35\times 10^{-13} \mbox{ GeV}\cong 9.6\times 10^{-32}M_{\rm pl}
$.

\end{enumerate}

\section{Gravitational reheating formulas}
\label{sec-2}

This section provides a detailed review of the results recently obtained in \cite{Haro:2024, Haro:2024a} (see also \cite{Hashiba:2018iff, Chun:2009yu,deHaro:2022ukj,Kaneta:2019a} to find some of the recent results in the context of gravitational reheating). 
To begin with we consider a potential which near the minimum $\varphi=0$ behaves like 
$\varphi^{2n}$ ($n$ is a natural number). We examine heavy massive $X$-particles, which are produced gravitationally
due to the coupling of a $X$-field with the Ricci scalar (see, for instance, \cite{Hashiba:2018iff},
and also \cite{Grib:1980aih,Parker:2009uva} for the foundations on  quantum field theory in curved spaces and its gravitational effects), 
and then they decay into Standard Model (SM) particles in order to reheat the universe.  During the oscillations, close to the minimum, as the potential behaves like $\varphi^{2n}$, hence, with the use of virial theorem, the effective Equation of State (EoS) parameter, $w_{\rm eff}$, becomes $w_{\rm eff}=\frac{n-1}{n+1}$. Note that $w_{\rm eff}$ is higher than $1/3$ for $n>2$. 
This guarantees that the inflaton's energy density decays faster than the energy density of the produced particles as well as their decay products, 
because, as a function of the scale factor, the energy density of matter decays as $a^{-3}$, the one of radiation as $a^{-4}$, and for the fluid with EoS parameter $w_{\rm eff}$,  it decays as $a^{-3(1+w_{\rm eff})}$, which in the case of the   inflaton's energy density   becomes $a^{-\frac{6n}{n+1}}$.\footnote{Without any loss of generality, we have set $a_0$, the present day value of the scale factor to be unity.}   As a consequence, the  energy density of the latter products will eventually dominate and finally this will lead to a successful reheating of the universe.    We shall concentrate on the class of inflationary potentials having  
similar behavior close to the minimum, e.g.  Hyperbolic Inflation, Superconformal $\alpha$-Attractor E-models  or Superconformal $\alpha$-Attractor T-models
\cite{Brown:2017osf,Linde:2013,Linde:2013a}.

{
In order to clearly realize the mechanism of gravitational reheating, it is essential to understand the decay process. The decay process can be described with the use of the dynamics governed by the Boltzmann equations:  

 \begin{eqnarray}\label{Boltzmann}
\left\{\begin{array}{ccc}
       \frac{d \rho_X(t)}{dt}+3H \rho_X(t)=-\Gamma_X  \rho_X(t)\\
       \bigskip \\
       \frac{d \rho_r(t)}{dt}+4H \rho_r(t)=\Gamma_X  \rho_X(t),
    \end{array}\right.
\end{eqnarray}
where $\rho_X(t)$ stands for the energy density of the produced $X$-particles,   $\rho_r(t)$ is the energy density of the radiation (it is the the energy density of the decay products), 
and $\Gamma_X$ is the decay rate of $X$-particles into SM particles and it is assumed to be a constant.
For example, 
considering an interaction of the $X$-field
with fermions,  this leads to the decay of the $X$-particles with the decay rate 
$\Gamma_X=\frac{\hat{h}^2m_X}{8\pi}$,  where $\hat{h}$ is a dimensionless constant \cite{Felder:1999pv}.

We proceed with the solution which describes the energy density of the heavy massive particles as follows
\begin{align}\label{energy-density-massive-particles}
\rho_X(t)=
    \rho_{X,\rm END}\left(
  \frac{a_{\rm END}}{a(t)}  \right)^3
  e^{-\Gamma_X(t-t_{\rm END})},  \quad t\geq t_{\rm END}.
\end{align}
Here, this solution  represents that the decay begins at the end of inflation, as discussed in \cite{Mambrini:2021zpp}. This is due to the fact that the heavy $X$-particles are produced at the end of inflation, when the inflaton starts to oscillate. Consequently, the decay of these particles commences immediately upon their creation.

Now, inserting (\ref{energy-density-massive-particles}) to the second equation of (\ref{Boltzmann}) and considering the fact that  decay starts
at the end of inflation, one can obtain
\begin{align}
    \rho_{ r}(t)=\rho_{X,\rm END}\left(
  \frac{a_{\rm END}}{a(t)}  \right)^4
  \int_{t_{\rm END}}^t \frac{a(s)}{a_{\rm END}}\Gamma_X
  e^{-\Gamma_X(s-t_{\rm END})}ds.
\end{align}
We now proceed by defining $t_{\rm dec}$ as the time where decay ends, that is,  when $\Gamma_X(t_{\rm dec}-t_{\rm END})\sim 1$. This quickly gives  $t_{\rm dec}\sim t_{\rm END}+\frac{1}{\Gamma_X}$, ($\Gamma_X \neq 0$), and
we focus on the investigation of the evolution of the decay products for $t \gg t_{\rm dec}$. 
We presume that background dominates and with such assumption  we have
$a(s)\cong a_{\rm END}\left(s/t_{\rm END} \right)^{\frac{n+1}{3n}}$ since $w_{\rm eff}=\frac{n-1}{n+1}$. Now, bearing in mind that
\begin{align}
& \int_{t_{\rm END}}^{\infty}
\left(s/t_{\rm END} \right)^{\frac{n+1}{3n}}\Gamma_X
e^{-\Gamma_X(s-t_{\rm END})}ds 
\nonumber\\  &
\cong 
\bar{\Theta}_X^{-\frac{n+1}{6n}}
\Gamma
\left( \frac{4n+1}{3n} \right)\cong 
\bar{\Theta}_X^{-\frac{n+1}{6n}}
,
\end{align}
where $\Gamma$ denotes the Euler's Gamma function, we  reach at the conclusion that for $t>t_{\rm dec}$, the following approximation can be made
\begin{eqnarray}
\rho_{ r}(t)\cong\rho_{X,\rm END}
\bar{\Theta}_X^{-\frac{n+1}{6n}}
\left(
  \frac{a_{\rm END}}{a(t)}  \right)^4,
\end{eqnarray}
where we have used the definition of the decay efficiency of $X$-particles.
Now since, $w_{\rm eff}=\frac{n-1}{n+1}$,
the energy density of the background,  that is,  the energy density of the inflaton field, 
satisfies the equation $\dot{\rho}_B=(1+w_{\rm eff})\rho_B$, and thus, 
it
 evolves as 
\begin{align}
    \rho_{\rm B}(t)=
    \rho_{\rm B, \rm END}
     \left( \frac{a_{\rm END}}{a(t)}\right)^{\frac{6n}{n+1}}.
    \end{align}
    
Therefore, since   
 the universe becomes reheated when
$\rho_{\rm B}\sim \rho_{r} $,  we get 
\begin{eqnarray}\label{relation}
    \Theta_X
= 
\bar{\Theta}_X^{\frac{n+1}{6n}}      
\left(\frac{a_{\rm END}}{a_{\rm reh}} \right)^
    {\frac{2(n-2)}{n+1}},
\end{eqnarray}
and as a result of which, the energy density of the decay products at the time of reheating is given by the following
\begin{eqnarray}
\label{energydensity}
   \rho_{ r, \rm reh}\cong
   \rho_{\rm B,\rm END}
   \bar{\Theta}_X^{-\frac{n+1}{2(n-2)}}   
   \Theta_X^{\frac{3n}{n-2}}.
\end{eqnarray}
Now, using the Stefan-Boltzmann law
 $T_{\rm reh}=\left( \frac{30}{\pi^2 g_{\rm reh}}\right)^{1/4}
 \rho_{ r, \rm reh}^{1/4}$  (here $g_{\rm reh}=106.75$ 
 denotes the effective number of degrees of freedom in the SM), 
 one gets the following reheating temperature,
\begin{eqnarray}
\label{reheating_temperature}
T_{\rm reh}
    =\left( \frac{90}{\pi^2 g_{\rm reh}}\right)^{1/4}
\bar{\Theta}_X^{-\frac{n+1}{8(n-2)}}
\Theta_X^{\frac{3n}{4(n-2)}}\sqrt{H_{\rm END}M_{\rm pl}}. 
\end{eqnarray}

It is essential to calculate the range of values for $\Gamma_X$. We first note that $\Gamma_X\ll H_{\rm END}$ because the decay occurs well after the end of inflation. Additionally, we have assumed that the energy density of the background dominates at the end of decay, that means, 
$ \rho_{ r,\rm  dec}\ll \rho_{\rm B, \rm dec}\cong 3M_{\rm pl}^2\Gamma^2_X$. Therefore, using the relation 
$\rho_{\rm B, \rm dec}\cong 3 M_{\rm pl}^2\Gamma^2_X$, we get,
\begin{eqnarray}
    \left(\frac{a_{\rm END}}{a_{\rm dec}} \right)^4\cong 
\bar{\Theta}_X^{\frac{2(n+1)}{3n}},    
\end{eqnarray}
and inserting the above relation to 
$\rho_{ r,\rm  dec}\ll  3M_{\rm pl}^2\Gamma^2_X$, one arrives at 
\begin{eqnarray}\label{Constraint}
    \Theta^{\frac{n}{n-1}}_X    \ll 
\sqrt{\bar{\Theta}_X}
\ll 
    1.
\end{eqnarray}

Moreover, combining   (\ref{Constraint})
 with the following bound of  the reheating temperature, 
$5\times 10^{-22} M_{\rm pl}\leq T_{\rm reh}\leq
5\times 10^{-10} M_{\rm pl}$, 
which states that
the reheating temperature
remains in an interval consistent with the Big Bang Nucleosynthesis (BBN) which occurs at about $\sim 1$ MeV scale,  and attains an upper bound at about $\sim 10^9$ GeV in order to mitigate the issues related to the gravitino problem ~\cite{Ellis:1982yb,Khlopov:1984pf,Kawasaki:2004qu,Kawasaki:2017bqm},
one gets four distinct cases.  
However,  the only viable case is the following 

\begin{eqnarray}
    \Theta_X^{\frac{n}{n-1}}\ll \sqrt{\bar{\Theta}_X}
    \leq 
     10^{\frac{84(n-2)}{n+1}}\left(
    \frac{H_{\rm END}}{M_{\rm pl}}
    \right)^{ \frac{2(n-2)}{n+1} }\Theta_X^{\frac{3n}{n+1}},
\end{eqnarray}

{
provided that the inequality
\begin{align}
     10^{-\frac{42(n-1)}{n}}\left(
    \frac{M_{\rm pl}}
    {H_{\rm END}}
    \right)^{ \frac{n-1}{n} }
    \ll
    \Theta_X\ll
     10^{-\frac{28(n-2)}{n}}\left(
    \frac{M_{\rm pl}}
    {H_{\rm END}}
    \right)^{ \frac{2(n-2)}{3n} },
\end{align}
holds. 
Finally, we comment on the maximum value of the reheating temperature. The maximum reheating temperature is obtained at the epoch when the decay coincides with the end of the inflaton's domination, that means, when $\rho_{ r,\rm reh}
\sim 3\Gamma^2_X M_{\rm pl}^2$. Now, considering eqn.  (\ref{energydensity}), one gets
\begin{eqnarray}
    \sqrt{\bar{\Theta}_X}
    \sim \Theta_X^{\frac{n}{n-1}},
\end{eqnarray}
and  inserting this into (\ref{reheating_temperature}),
maximum reheating temperature becomes

\begin{eqnarray}\label{Tem_max}
    T_{\rm reh}^{\rm max}
\cong 5.4\times 10^{-1}
\Theta_X^{\frac{n}{2(n-1)}}\sqrt{H_{\rm END}M_{\rm pl}},\end{eqnarray}
where $\Theta_X$, due to the bounds of the reheating temperature,  is constrained as follows
\begin{align}
     10^{-\frac{42(n-1)}{n}}\left(
    \frac{M_{\rm pl}}
    {H_{\rm END}}
    \right)^{ \frac{n-1}{n} }
    \leq
    \Theta_X\leq
     10^{-\frac{18(n-1)}{n}}\left(
    \frac{M_{\rm pl}}
    {H_{\rm END}}
    \right)^{ \frac{n-1}{n} }.
\end{align}

\section{Gravitational production of dark matter}
\label{sec-3}

This section deals with the gravitational production of dark matter in the framework of gravitational reheating. We consider two quantum scalar fields, $X$ and $Y$ which are conformally coupled to the Ricci scalar. The $X$-field produces heavy 
$X$-particles with mass $m_X$, which will decay into SM particles and reheat the universe. 
The $Y$-field produces $Y$-particles with 
 mass $m_Y$, and they correspond to the present-day dark matter candidate.

\subsection{Maximum reheating temperature}
\label{max_reh_tem}

The case when the $X$-particles decay close to the onset of radiation leads to the maximum reheating temperature. 
We begin by considering the energy density of the dark matter at the present time which is given by 
\begin{eqnarray}\label{Y0}
    \rho_{Y,0}=
    \rho_{Y,\rm END}\left(\frac{a_{\rm END}}{a_0} \right)^3=
    \Theta_Y \rho_{\rm B,\rm END}\left(\frac{a_{\rm END}}{a_0} \right)^3.\end{eqnarray}
Now, with the use of the following
\begin{eqnarray}\label{Y1}
\left(\frac{a_{\rm END}}{a_0} \right)^3=
\left(\frac{a_{\rm END}}{a_{\rm reh}} \right)^3
\left(\frac{a_{\rm reh}}{a_0} \right)^4\frac{a_0}{a_{\rm reh}}
\nonumber\\
=
\left(\frac{a_{\rm END}}{a_{\rm reh}} \right)^3
\left(\frac{a_{\rm reh}}{a_0} \right)^4
\frac{T_{\rm reh}^{\rm max}}{T_0},\end{eqnarray}
where the adiabatic expansion of the universe after reheating has been considered, i.e. $a_0T_0=a_{\rm reh}T_{\rm reh}^{\rm max}$, we now calculate
\begin{eqnarray}\label{Y2}
&&\rho_{r,0}\left(\frac{a_{0}}{a_{\rm reh}} \right)^4
\left(\frac{a_{\rm reh}}{a_{\rm END}} \right)^3   
\nonumber\\&& 
=
    \rho_{\rm B,\rm reh}
    \left(\frac{a_{\rm reh}}{a_{\rm END}} \right)^3=
    \rho_{\rm B,\rm END}\Theta_X
    ,\end{eqnarray}
wherein we make use of 
$\left(\frac{a_{\rm END}}{a_{\rm reh}} \right)^3
=\Theta_X^{\frac{n+1}{n-1}}$ 
\cite{Haro:2024a}, 
which arises under the assumptions of the background  evolves as 
\begin{align}
    \rho_{\rm B}(t)=
    \rho_{\rm B,\rm END}\left(a_{\rm END}/a(t)\right)^{\frac{6n}{n+1}},
    \end{align}
    and the decay of the $X$-particles is at the onset of radiation. With these,  we find the following relation between the energy density of the Y-particles and the maximum reheating temperature
\begin{eqnarray}
    \rho_{Y,0}=
    \frac{\Theta_Y}{\Theta_X   }\rho_{r,0}\frac{T_{\rm reh}^{\rm max}}{T_0},\end{eqnarray}
    that is:
    \begin{eqnarray}
\Omega_Yh^2=\Omega_rh^2\frac{\Theta_Y}{\Theta_X }
\frac{T_{\rm reh}^{\rm max}}{T_0}.    \end{eqnarray}
Now, inserting the value of the 
maximum reheating temperature and the observational values of $\Omega_Y h^2$ and $\Omega_{r}h^2$, one arrives at the following bound 
\begin{eqnarray}
8.63\times 10^{-28}\sqrt{\frac{M_{\rm pl}}{H_{\rm END}}}=\Theta_Y\Theta_X^{\frac{2-n}{2(n-1)}}  
\end{eqnarray}

Next, we deal with scalar particles conformally coupled to gravity (for the non-conformally coupled case, specially the minimally coupled case,  see for instance \cite{Kolb:2023ydq,Jenks:2024fiu}), when  $m_A \ll H_{\rm END}$,  because in this case one can use  the Wentzel-Kramers-Brillouin (WKB) method in the complex plane in order to analytically calculate the $\beta$-Bogoliubov coefficients, which is the key piece to find the energy density of the produced particles.
Using the results obtained in \cite{Haro:2024a}, 
we get 
\begin{align}\label{Theta}
    \Theta_A=
    \frac{1}
{12\pi^3}\left(\frac{m_A}{M_{\rm pl}}\right)^{5/2}\sqrt{\frac{M_{\rm pl}}{\sqrt{2}H_{\rm END}}}\cong 2.26
\left(\frac{m_A}{M_{\rm pl}}\right)^{5/2},
\end{align}
where we have used the usual value of the Hubble rate at the end of inflation, i.e. $H_{\rm END}\sim 10^{-6}M_{\rm pl}$.
Then, we obtain:
\begin{eqnarray}\label{mx-my-eqn}
    && \frac{m_Y}{M_{\rm pl}}\cong     1.7\times 10^{-10}\Theta_X^{\frac{n-2}{5(n-1)}}\nonumber\\&&
    \cong  1.7\times 10^{-10}\left(
    2.26
\left(\frac{m_X}{M_{\rm pl}}\right)^{5/2}     \right)^{\frac{n-2}{5(n-1)}},
\end{eqnarray}
and taking into account the constraint, 
$5\times 10^{-22}M_{\rm pl}\leq T_{\rm reh}^{\rm max}\leq 
5\times 10^{-10}M_{\rm pl}$, coming from the BBN success,  after substituting (\ref{Tem_max}) into it, we get:
\begin{eqnarray}
& 10^{-21}\sqrt{\frac{M_{\rm pl}}{H_{\rm END}}}\leq
    \Theta_X^{\frac{n}{2(n-1)}}
    \leq 10^{-9}\sqrt{\frac{M_{\rm pl}}{H_{\rm END}}} \nonumber\\ \Longrightarrow &
     10^{-18} \leq
    \Theta_X^{\frac{n}{2(n-1)}}
    \leq 10^{-6},  
    \end{eqnarray}}
    that is:
    \begin{eqnarray}
    0.7\times 10^{-\frac{72(n-1)}{5n}}\leq \frac{m_X}{M_{\rm pl}}\ll 10^{-6},    
    \end{eqnarray}
    where we have used that $m_X\ll H_{\rm END}\cong 10^{-6}M_{\rm pl}$. 
    Finally, taking into account (\ref{Tem_max}) and (\ref{Theta}), the reheating temperature can be written as a function of the mass $m_X$ and the parameter $n$,  as follows:
    \begin{eqnarray}
    T_{\rm reh}^{\rm max}=5.4\times 10^{-4}\left(2.26 \left(\frac{m_X}{M_{\rm pl}}\right)^{5/2}   \right)^{\frac{n}{2(n-1)}}
    M_{\rm pl}.
    \end{eqnarray}

\begin{figure*}
\centering
\includegraphics[width=0.7\textwidth]{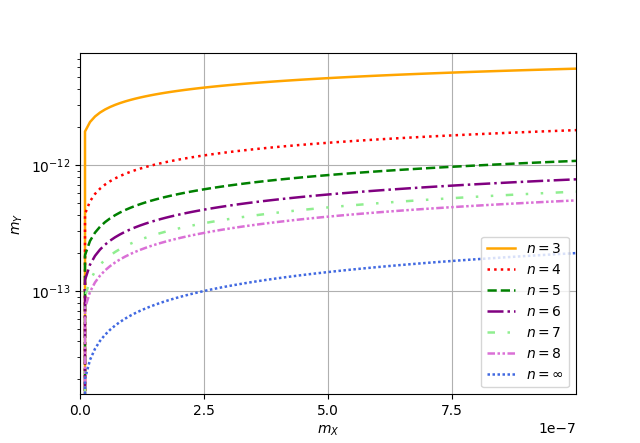}
\caption{The dependence of $m_Y$ masses with $m_X$ has been shown for different values of $n$. We work with the units where $M_{\rm pl} =1$. }
\label{fig:1}
\end{figure*}
\begin{figure*}
\centering
\includegraphics[width=0.7\textwidth]{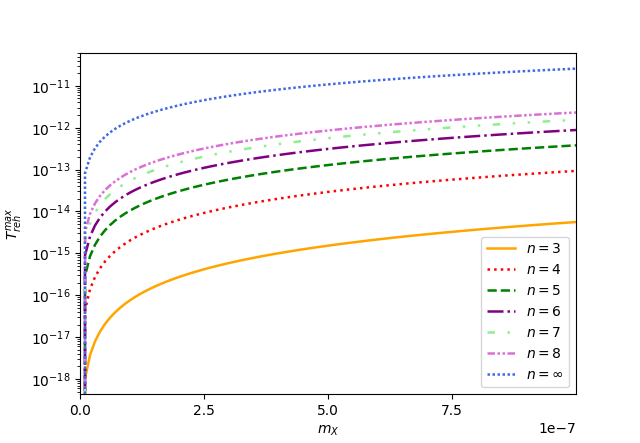}
\caption{The maximum reheating temperature versus the mass of the produced particles, $m_X$ for different values of $n$ has been depicted. We work with the units where $M_{\rm pl} =1$. }
\label{fig:2}
\end{figure*}

Finally, we close this section with figures \ref{fig:1} and \ref{fig:2}. In Fig. \ref{fig:1} we display the dependence of $m_Y$ with $m_X$ as given in eqn. (\ref{mx-my-eqn}) for different values of $n$. This clearly exhibits a pattern for increasing $n$. On the other hand, in Fig. \ref{fig:2} we show how the maximum reheating temperature, $T^{\rm max}_{\rm reh}$, depends on $m_X$ for different values of $n$. One can clearly notice that for a specific value of $n$, if $m_X$ increases,  $T^{\rm max}_{\rm reh}$ also increases. Additionally, one can further notice that for a particular value of $m_X$, if $n$ increases, then $T^{\rm max}_{\rm reh}$ increases and it assumes the maximum value for $n = \infty$.

\subsubsection{Quintessential Inflation}

In this section we discuss the bounds on the masses of $X$ and $Y$ particles in a special cosmological scenario, namely, 
Quintessential Inflation 
\cite{Peebles:1998qn,Giovannini:1999bh,Dimopoulos:2001ix,Giovannini:2003jw,Sami:2004xk,Rosenfeld:2005mt,Chun:2009yu, Bento:2009zz,Lankinen:2016ile,DeHaro:2017abf,AresteSalo:2017lkv,Hashiba:2018iff, Haro:2018zdb, deHaro:2019oki,
deHaro:2022ukj,Dimopoulos:2022rdp,deHaro:2023gho,Inagaki:2023mxv,Giare:2024sdl} -- a unified cosmological  model where after inflation the universe enters in a kination phase, i.e., all the energy density is kinetic which means that $w_{\rm eff}=1$, and thus, it is equivalent to the case $n=\infty$. 
Using the values of $\Theta_X$ and $\Theta_Y$, we have the following relation between the masses
\begin{eqnarray}
    \frac{m_Y}{M_{\rm pl}}\cong
    5.64\times 10^{-11}\left(\frac{M_{\rm pl}}{H_{\rm END}}
\right)^{1/10}\sqrt{\frac{m_X}{M_{\rm pl}}},
\end{eqnarray}
which for $H_{\rm END}\cong 10^{-6}M_{\rm pl}$, leads to 
\begin{eqnarray}
    \frac{m_Y}{M_{\rm pl}}\cong
    2\times 10^{-10}
\sqrt{\frac{m_X}{M_{\rm pl}}}.
\end{eqnarray}
On the other hand, the bound of the maximum reheating temperature leads to the following constraint
\begin{eqnarray}
    10^{-18}\leq
    \sqrt{\Theta_X}
    \leq 10^{-6},\end{eqnarray}
where we continue taking 
$H_{\rm END}\cong 10^{-6}M_{\rm pl}$. Inserting the value of $\Theta_X\cong 2.26\left(\frac{m_X}{M_{\rm pl}}\right)^{5/2}$, we arrive at the following 
\begin{eqnarray}
    5.51\times 10^{-15}\leq 
    \frac{m_X}{M_{\rm pl}}
    \leq 2.2\times 10^{-5},
    \end{eqnarray}
which must be improved due to the fact that we have assumed $m_X\ll H_{\rm END}\cong 10^{-6}M_{\rm pl}$, leading to, 
\begin{eqnarray}
    5.51\times 10^{-15}\leq 
    \frac{m_X}{M_{\rm pl}}
    \ll  10^{-6}.
    \end{eqnarray}

Now, using this last constraint, we find the range of viable values of the mass of dark matter lying in the following region:
\begin{eqnarray}
    1.48\times 10^{-17}\leq 
    \frac{m_Y}{M_{\rm pl}}
    \ll 10^{-13}.
    \end{eqnarray}

\subsection{General case: decay before the onset of radiation}
When the decay is before the end of the inflaton's domination, using (\ref{relation}) and (\ref{energydensity}), one has, 
\begin{align}
    \rho_{\rm B,\rm reh}\left(
    \frac{a_{\rm reh}}{a_{\rm END}}
    \right)^3=\rho_{\rm B,\rm END}\Theta_X^{\frac{3(n-1)}{2(n-2)}} \bar{\Theta}_X^{-\frac{n^2-1}{4n(n-2)}}    
    .
    \end{align}
Therefore,  from (\ref{Y0}), (\ref{Y1}) and (\ref{Y2}),  the energy density of the $Y$ particles  becomes 
\begin{eqnarray}
    \rho_{Y,0}=\Theta_Y
    \Theta_X^{-\frac{3(n-1)}{2(n-2)}}
\bar{\Theta}_X^{\frac{n^2-1}{4n(n-2)}}     \rho_{r,0} \frac{T_{\rm reh}}{T_0},  \end{eqnarray}
and taking into consideration  the formula of the reheating temperature (\ref{reheating_temperature}) one arrives at,
\begin{align}
    \rho_{Y,0}=
    \left(\frac{90}{\pi^2 g_{\rm reh}} \right)^{1/4}
    \Theta_Y
    \Theta_X^{-3/4}
\bar{\Theta}_X^{\frac{n+1}{8n}}    
    \rho_{r,0} \frac{\sqrt{H_{\rm END}M_{\rm pl}}}{T_0},  \end{align}
    which in terms of the density parameters, takes the form 
\begin{align}
    \Omega_{Y}h^2=
    \left(\frac{90}{\pi^2 g_{\rm reh}} \right)^{1/4}
    \Theta_Y
    \Theta_X^{-3/4}
\bar{\Theta}_X^{\frac{n+1}{8n}}    
    \Omega_{r}h^2 \frac{\sqrt{H_{\rm END}M_{\rm pl}}}{T_0}. \end{align}    
    This coincides with the previous case when 
    $\sqrt{\bar{\Theta}_X}
    =\Theta_X^{\frac{n}{n-1}}$. Now, inserting the observational values, one gets,
\begin{eqnarray}
    8.67\times 10^{-28}\sqrt{\frac{M_{\rm pl}}{H_{\rm END}}}=\Theta_Y
    \Theta_X^{-3/4}
\bar{\Theta}_X^{\frac{n+1}{8n}},    
    \end{eqnarray}
and from the expression of $\Theta_Y$, we derive
\begin{eqnarray}
   \frac{m_Y}{M_{\rm pl}}\cong
   1.7\times 10^{-10}\Theta_X^{3/10}
   \bar{\Theta}_X^{-\frac{n+1}{20n}}
   ,\end{eqnarray}
   with the following constraints (see for details \cite{Haro:2024a}):
\begin{eqnarray}
     10^{-\frac{36(n-1)}{n}}
    \ll
    \Theta_X\ll
     10^{-\frac{24(n-2)}{n}}
    ,
\end{eqnarray}
and 
\begin{eqnarray}
    \Theta_X^{\frac{n}{n-1}}\ll
    \sqrt{\bar{\Theta}_X}
    \leq 
     10^{\frac{72(n-2)}{n+1}}
    \Theta_X^{\frac{3n}{n+1}},
\end{eqnarray}
where we continue using 
$H_{\rm END}\cong 10^{-6} M_{\rm pl}$. 
From this last constraint, it
 is also possible to obtain the bound for $m_Y$ as, 
\begin{eqnarray}
    1.7\times 10^{-10}
    10^{-\frac{36(n-2)}{5n}}
    \leq \frac{m_Y}{M_{\rm pl}}\ll 2\times 10^{-10}\Theta_X^{\frac{n-2}{5(n-1)}},
\end{eqnarray}
with $\Theta_X\cong 2.26\left(\frac{m_X}{M_{\rm pl}}\right)^{5/2}$ and 
\begin{eqnarray}
    10^{-\frac{72(n-1)}{5n}}
    \ll
    \frac{m_X}{M_{\rm pl}}
    \ll 
     \mbox{min}(10^{-\frac{48(n-2)}{5n}}; 10^{-6}).
\end{eqnarray}

\subsubsection{Quintessential Inflation}

In the case with $n=\infty$, the formulas get simplifies as follows. 
The reheating temperature in this case is given by
\begin{align}
\label{reheating_temperature_quintessential}
T_{\rm reh}
    & =\left( \frac{90}{\pi^2 g_{\rm reh}}\right)^{1/4}
\bar{\Theta}_X^{-\frac{1}{8}}
\Theta_X^{\frac{3}{4}}\sqrt{H_{\rm END}M_{\rm pl}}
\nonumber \\ 
&\cong 5.4 \times 10^{-4}
\bar{\Theta}_X^{-\frac{1}{8}}
\Theta_X^{\frac{3}{4}}\sqrt{H_{\rm END}M_{\rm pl}}. 
\end{align}
The relation between different heating efficiencies and the  decay efficiency is given by 
\begin{eqnarray}
    8.67\times 10^{-25}=\Theta_Y
    \Theta_X^{-3/4}
\bar{\Theta}_X^{\frac{1}{8}},    
    \end{eqnarray}
which leads to 
\begin{eqnarray}
   \frac{m_Y}{M_{\rm pl}}\cong
   1.7\times 10^{-10}\Theta_X^{3/10}
   \bar{\Theta}_X^{-\frac{1}{20}}
   ,\end{eqnarray}
with the constraints:
\begin{eqnarray}
     10^{-36}
    \ll
    \Theta_X\ll
     10^{-{24}},
\end{eqnarray}
and 
\begin{eqnarray}
    \Theta_X\ll
    \sqrt{\bar{\Theta}_X}
    \leq 
     10^{{72}}
    \Theta_X^{{3}}.
\end{eqnarray}  
Finally, we have the following bound for the masses:
\begin{eqnarray}
    4\times 10^{-14}
    \ll
    \frac{m_X}{M_{\rm pl}}
    \ll  3\times
     10^{-10},
\end{eqnarray}
and 
\begin{eqnarray}
    1.07\times 10^{-17}
    \leq \frac{m_Y}{M_{\rm pl}}\ll 2\times 10^{-10}\Theta_X^{\frac{1}{5}},
\end{eqnarray}
with $\Theta_X\cong 2.26\left(\frac{m_X}{M_{\rm pl}}\right)^{5/2}$.
Taking into account the bound of $\Theta_X$, one can conclude that the viable masses for dark matter are of the 
order $(10^{-17}-10^{-16}) M_{\rm pl}\sim (10^1-10^2)$ GeV.

\section{Gravitational production of dark matter + reheating via inflaton decay}
\label{sec-4}

As we have already discussed, gravitational reheating works only for $n\geq 3$. 
The reason behind this is that the energy density of the background must decrease faster than that of the decay products for radiation to eventually dominate and for the universe to become reheated. 
As we have already discussed, 
this condition is satisfied only when, after inflation, the effective EoS parameter of the background is greater than
 $1/3$, which corresponds to  $n\geq 3$. 

Here we investigate  the case with $n=1$, 
where the potential behaves like $\varphi^2$ near its minimum, corresponding to models such as Higgs or Starobinsky inflation. 
To achieve a successful reheating, we assume that the inflaton field decays into SM particles with a decay rate 
$\Gamma_{\varphi}$. 
In this scenario, we first assume that the decay is instantaneous, that means, all the energy stored in the inflaton field is immediately converted to radiation, which is quite different than a delayed decay as we will see in section  \ref{delayed_decay}. Reheating is achieved when
 $ \Gamma_{\varphi}\cong H_{\rm reh}$, yielding  a reheating temperature 
 given by \cite{Kofman:1997yn}:
\begin{align}
    T_{\rm reh}\cong 5.4\times 10^{-1}\sqrt{\Gamma_{\varphi}M_{\rm pl}}. 
\end{align}
Next, we consider the scalar $Y$-particles, which are created gravitationally and are responsible for the present dark matter. 
For $n=1$, the background evolves as
$\rho_{\rm B}(t)=\rho_{\rm B,\rm END}\left( \frac{a_{\rm END}}{a(t)}\right)^3$, and immediately after the decay which coincides with the reheating time, this energy density transforms to 
$\rho_r(t)=\rho_{r,\rm reh}\left(\frac{a_{\rm reh}}{a(t)}\right)^4$. Then, 
following step by step the calculation done in section~\ref{max_reh_tem}, we obtain,
\begin{eqnarray}
\Omega_Yh^2=\Omega_rh^2{\Theta_Y}
\frac{T_{\rm reh}}{T_0},    \end{eqnarray}
which coincides with the result used in 	\cite{Chung:2001cb}. 
By inserting the value of the reheating temperature along with the observational data, we obtain:
\begin{eqnarray}
\Theta_Y\cong 8.63\times 10^{-28}\sqrt{\frac{M_{\rm pl}}{\Gamma_{\varphi}}}.
\end{eqnarray}
At this point, we note that for
$m_Y\ll H_{\rm END}$, when
$n=1$, we have \cite{Haro:2024a, Ema:2018ucl, Kolb:2023ydq}
\begin{align}
\rho_{Y,\rm END}\cong \frac{9}{4\pi}\Gamma^{-4}\left(\frac{1}{4}\right)m_Y^2H_{\rm END}^2
    \cong 4.14\times 10^{-3}
    m_Y^2H_{\rm END}^2,
\end{align}
where we have used $
\Gamma\left(\frac{1}{4}\right)\cong 3.6254$. Therefore, $\Theta_Y\cong 1.38\times 10^{-3}\left(\frac{m_Y}{M_{\rm pl}}\right)^2$, which leads to the following relation between $m_Y$ and $\Gamma_{\varphi}$ as 
\begin{align}
\left(\frac{m_Y}{M_{\rm pl}}\right)^2\cong 6.25\times 10^{-25}\sqrt{\frac{M_{\rm pl}}{\Gamma_{\varphi}}}.
\end{align}

Finally, we derive the relation between the reheating temperature and the dark matter mass which goes as 
\begin{align}
    T_{\rm reh}\cong 3.37\times 10^{-25}\left(\frac{M_{\rm pl}}{m_Y}    \right)^2
    M_{\rm pl},\end{align}
and taking this into account, the constraint of the reheating temperature we find to be,
\begin{eqnarray}
\label{masses}
    2.59\times 10^{-8}\leq \frac{m_Y}{M_{\rm pl}}\ll 
    10^{-6},
\end{eqnarray}
where we have taken into account that $m_Y\ll H_{\rm END}\sim 10^{-6}M_{\rm pl}$.
Therefore, the viable masses for dark matter are of the order: $m_Y\sim 10^{-7}M_{\rm pl}\sim 2\times 10^{11}$ GeV.

\subsection{Delayed decay}\label{delayed_decay}

This situation is more involved than the previous case because it requires to solve the Boltzmann equations:
\begin{eqnarray}\label{Boltzmann_1}
\left\{\begin{array}{ccc}
       \frac{d  \rho_{\rm B}(t)}{dt}+3H \rho_{\rm B}(t)=-\Gamma_{\varphi}\rho_{\rm B}(t),\\
       \bigskip \\
       \frac{d \rho_r(t)}{dt}+4H \rho_r(t)=\Gamma_{\varphi}  \rho_{\rm B}(t),
    \end{array}\right.
\end{eqnarray}
where for simplicity, we assume that the decay rate is constant.
Following the reasoning depicted in \cite{Mambrini:2021zpp}
 we can write the second equation as follows
\begin{align}
     \dfrac{d(\rho_r a^3)}{da}=\dfrac{\Gamma_{\varphi}\rho_{\rm B}}{H}a^3,
\end{align}
and assuming that the decay occurs during the inflaton's domination, i.e., when $H\cong \sqrt{\frac{\rho_{\rm B}}{3M_{\rm pl}^2}}$, one has 
\begin{align}\label{rho_rad}
     \dfrac{d(\rho_r a^3)}{da}=
     \sqrt{3\rho_{\rm B}}
     \Gamma_{\varphi} M_{\rm pl}a^3.
\end{align}

Next, following \cite{Mambrini:2021zpp}, we also assume that the background scales as
$\rho_{\rm B}=\rho_{\rm B,\rm END}\left(a_{\rm END}/a \right)^3$, and inserting it into (\ref{rho_rad}), which after integration leads to 
\begin{align}
\rho_r=\frac{2\sqrt{3}}{5}
\Gamma_{\varphi}M_{\rm pl}
    \sqrt{\rho_{\rm B,\rm END}}
    \left(\frac{a_{\rm END}}{a} \right)^{3/2}
    \left[1- \left(\frac{a_{\rm END}}{a} \right)^{5/2}    \right],\end{align}
where we can see that for $a\gg a_{\rm END}$, the energy density of radiation evolves as
\begin{align}
\rho_r=\frac{2\sqrt{3}}{5}
\Gamma_{\varphi}M_{\rm pl}
    \sqrt{\rho_{\rm B,\rm END}}
    \left(\frac{a_{\rm END}}{a} \right)^{3/2}
   ,\end{align}
which does not match with the real evolution of radiation at late times, which scales as $a^{-4}$.  
The problem in this approach is that the authors assume, 
for all the time, an  evolution of the background which only holds at early times when $\Gamma_{\varphi}\ll H$ as is discussed in the earlier works~\cite{Garcia:2020eof,
 Chung:2001cb,
 Kaneta:2019a,
 Giudice:2001},
because as we can see from the first equation of (\ref{Boltzmann_1}), its exact evolution, for all time $t\geq t_{\rm END}$,  is 
\begin{align}
    \rho_{\rm B}(t)=\rho_{\rm B,\rm END}\left(\frac{a_{\rm END}}{a(t)} \right)^3
    e^{-\Gamma_{\varphi}(t-t_{\rm END})},\end{align}
which means that at very late times, since the right hand side of the second  equation of (\ref{Boltzmann_1})
 decays exponentially,  it becomes 
$\frac{d \rho_r(t)}{dt}+4H \rho_r(t)=0$, which is the conservation equation for radiation. 
Therefore, as we will demonstrate, the reheating temperature obtained in this work will differ slightly—by less than an order of magnitude—from the values reported in previous studies. 

Having clarified this point, we now address the exact evolution of radiation, which is described by:
\begin{align}\label{rho_reheating1}
    \rho_{r}(t)=\rho_{\rm B,\rm END}\left(
  \frac{a_{\rm END}}{a(t)}  \right)^4
  \int_{t_{\rm END}}^t \frac{a(s)}{a_{\rm END}}\Gamma_{\varphi}
  e^{-\Gamma_{\varphi}(s-t_{\rm END})}ds.
\end{align}

Equating both energy densities at the reheating time, one has,
\begin{align}\label{reh_rel}
e^{-\Gamma_{\varphi}t_{\rm reh}}
=  \int_{t_{\rm END}}^{t_{\rm reh}} \frac{a(s)}{a_{\rm reh}}\Gamma_{\varphi}
  e^{-\Gamma_{\varphi}s}ds,\end{align}
which  is impossible to solve analytically,
because to find the evolution of the scale factor one needs to analytically solve (\ref{Boltzmann_1}) with 
$H(t)=\frac{\sqrt{\rho_{\rm B}(t)+\rho_r(t)}}{\sqrt{3}M_{\rm pl}}$, which can only be solved numerically.
Therefore, to find an approximate solution of (\ref{reh_rel}), we follow the strategy to assume that before the reheating the evolution is dominated by the background, and thus, the evolution of the scale factor is given by:
\begin{eqnarray}
    \frac{\dot{a}(t)}{a(t)}\cong
    H_{\rm END}\left(
  \frac{a_{\rm END}}{a(t)}  \right)^{3/2}
  e^{-\frac{\Gamma_{\varphi}}{2}(t-t_{\rm END})},\end{eqnarray}
which leads to 
\begin{align}
    \frac{a(t)}{a_{\rm END}}\cong
    \left(\frac{3H_{\rm END}}{\Gamma_{\varphi}} \right)^{2/3}\left[ 1-e^{-\frac{\Gamma_{\varphi}}{2}(t-t_{\rm END})}
    +\frac{\Gamma_{\varphi}}{3H_{\rm END}}
    \right]^{2/3},
\end{align}
which under the assumption of $\Gamma_{\varphi}\ll H_{\rm END}$
can be approximated by
\begin{align}\label{scale_factor}
    \frac{a(t)}{a_{\rm END}}\cong
    \left(\frac{3H_{\rm END}}{\Gamma_{\varphi}} \right)^{2/3}\left[ 1-e^{-\frac{\Gamma_{\varphi}}{2}t}
    \right]^{2/3}.
\end{align}

Therefore, 
\begin{align}
& \int_{t_{\rm END}}^t \frac{a(s)}{a_{\rm END}}\Gamma_{\varphi}
  e^{-\Gamma_{\varphi}(s-t_{\rm END})}ds
  \nonumber\\ &
  \cong \left( \frac{3H_{\rm END}}{\Gamma_{\varphi}}\right)^{2/3}
\int_0^{t\Gamma_{\varphi}}
  \left(1-e^{-x/2} \right)^{2/3}
  e^{-x}dx,
\end{align}
where we have assumed that $t_{\rm END}\Gamma_{\varphi}
\sim \Gamma_{\varphi}/H_{\rm END}\ll 1$.
Taking into account that a primitive, which can be obtained integrating by parts,  of the integrand is given by:
\begin{eqnarray}
    \frac{3}{20}\left(1-e^{-x/2} \right)^{5/3}
    \left(3+5e^{-x/2}\right),\end{eqnarray}
we have 
\begin{align}\label{radiation_formula}
&& \rho_{ r}(t)\cong
\frac{3}{20}\rho_{\rm B,\rm END}
\left( \frac{3H_{\rm END}}{\Gamma_{\varphi}}\right)^{2/3}
\left(1-e^{-t\Gamma_{\varphi}/2} \right)^{5/3}
\nonumber\\ &&
  \times 
   \left(3+5e^{-t\Gamma_{\varphi}/2}\right)
\left(
  \frac{a_{\rm END}}{a(t)}  \right)^4.
\end{align}
For $t>t_{\rm reh}$ the contribution of the term
\begin{align}
    \int_{t_{\rm reh}}^t \frac{a(s)}{a_{\rm END}}\Gamma_{\varphi}e^{-\Gamma_{\varphi}(s-t_{\rm END})}ds
\end{align}
appearing in (\ref{rho_reheating1}) can be disregarded, which means that, at late times $\rho_r(t)\cong \rho_{r,\rm reh}\left(\frac{a_{\rm reh}}{a(t)}\right)^4$.

Returning to (\ref{reh_rel}),
(\ref{scale_factor})  and   (\ref{radiation_formula}),   after introducing the notation,
$z_{\rm reh}\equiv e^{-t_{\rm reh}\Gamma_{\varphi}/2}$, we find:
\begin{eqnarray}
    35z_{\rm reh}^2-6z_{\rm reh}-9=0,
\end{eqnarray}
whose solution is $z_{\rm reh}=3/5$, and leads to the following reheating temperature
$\rho_{r,\rm reh}\cong \frac{3}{4}
\Gamma_{\varphi}^2
M_{\rm pl}^2
$, which is of the same order, the difference is to replace $3/4$ by $1/2$,  than the result obtained in \cite{Mambrini:2021zpp}
 using the method explained above. 
Having in mind all of these results, we calculate: 
\begin{eqnarray}
    \rho_{Y,0}=
    \rho_{Y,\rm END}\left(\frac{a_{\rm END}}{a_0} \right)^3=
    \Theta_Y \rho_{\rm B,\rm END}\left(\frac{a_{\rm END}}{a_0} \right)^3,\end{eqnarray}
and write
\begin{eqnarray}
\left(\frac{a_{\rm END}}{a_0} \right)^3 =
\left(\frac{a_{\rm END}}{a_{\rm reh}} \right)^3
\left(\frac{a_{\rm reh}}{a_0} \right)^4\frac{a_0}{a_{\rm reh}}
\nonumber\\
=
\left(\frac{a_{\rm END}}{a_{\rm reh}} \right)^3
\left(\frac{a_{\rm reh}}{a_0} \right)^4
\frac{T_{\rm reh}}{T_0},\end{eqnarray}
obtaining:
\begin{align}
\rho_{Y,0}=
    \Theta_Y \rho_{\rm B,\rm END}\left(\frac{a_{\rm END}}{a_{\rm reh}} \right)^3
\left(\frac{a_{\rm reh}}{a_0} \right)^4
\frac{T_{\rm reh}}
{T_0}.\end{align}
Now, multiplying the right hand side by $\dfrac{\rho_{r,0}}{\rho_{r,0}}$, we find, 
\begin{align}
& \rho_{Y,0}=
    \Theta_Y \rho_{\rm B,\rm END}\rho_{r,0}\frac{T_{\rm reh}}
{T_0} 
\frac{1}{\rho_{r,0} 
\left(\frac{a_0}{a_{\rm reh}}\right)^4
\left(\frac{a_{\rm reh}} {a_{\rm END}}\right)^3}\nonumber\\
& \cong \Theta_Y \rho_{\rm B,\rm END}\rho_{r,0}\frac{T_{\rm reh}}
{T_0} 
\frac{1}{\rho_{r,\rm reh} 
\left(\frac{a_{\rm reh}} {a_{\rm END}}\right)^3}
\frac{25}{9}
\Theta_Y \rho_{r,0}\frac{T_{\rm reh}}
{T_0},\end{align}
where we have used that 
$\rho_{r,\rm reh} 
\left(a_{\rm reh}/a_{\rm END}\right)^3\cong 
\frac{9}{25}\rho_{\rm B, \rm END}$.
Therefore, we conclude that the relation between the density parameters is,
\begin{eqnarray}
    \Omega_Y h^2\cong \frac{25}{9}
\Theta_Y \Omega_{r}h^2\frac{T_{\rm reh}}
{T_0},\end{eqnarray}
and inserting the observational data, we get,
\begin{align}
1.68\times 10^{-28}\cong 
    \Theta_Y\frac{T_{\rm reh}}{M_{\rm pl}},
\end{align}
and using that $\Theta_Y\cong 1.38\times 10^{-3}\left(\frac{m_Y}{M_{\rm pl}}\right)^2$, we find,
\begin{eqnarray}
    T_{\rm reh}\cong 1.22\times 10^{-25}
    \left(\frac{M_{\rm pl}}
    {m_Y}    \right)^2 M_{\rm pl},\end{eqnarray}
which is of the same order as the result obtained in the previous case of instantaneous decay. Consequently, the viable masses of dark matter are of the same order as those obtained in (\ref{masses}).

\section{Conclusions}
\label{sec-5-conclusion}

In this article, we have studied the gravitational production of  dark matter particles resulting from the conformal coupling of a massive scalar field to the Ricci tensor in two distinct scenarios: {\bf (i)} when the reheating occurs via gravitational production of heavy particles conformally coupled to gravity, which subsequently decay into SM particles, and {\bf (ii)} when reheating is driven by the decay of the inflaton field into SM particles.

In both scenarios, we relate the reheating temperature to the mass of dark matter particles. By considering the constraints on the reheating temperature imposed by the success of the BBN, we identify the range of viable masses for dark matter particles,  obtaining new bounds (see sections \ref{sec-3} and \ref{sec-4}).

A significant result emerges when reheating is driven by the gravitational production of scalar particles conformally coupled to gravity. In this case, as described in section \ref{sec-3},  the range of viable dark matter masses is relatively low and narrow, depending on the effective EoS parameter during the  oscillations of the inflaton field. For example, in the case of Quintessential Inflation, where $w_{\rm eff}=1$, the maximum viable mass is below the  TeV scale.

Conversely, in the second scenario where the potential is approximately quadratic near its minimum, as we have shown in section \ref{sec-4}, the viable mass range for dark matter particles is significantly higher, centered around $10^{11}$ GeV, though still within a narrow range. This highlights the strong dependence of viable dark matter properties on the specific dynamics of reheating in the early universe.

\section*{Acknowledgments} 
JdH is supported by the Spanish grants 
PID2021-123903NB-I00 and 
RED2022-134784-T
funded by MCIN/AEI/10.13039/501100011033 and by ERDF ``A way of making Europe''.
SP has been supported by the Department of Science and Technology (DST), Govt. of India under the Scheme   ``Fund for Improvement of S\&T Infrastructure (FIST)'' (File No. SR/FST/MS-I/2019/41).   

\bibliography{references}

\end{document}